\documentclass[10pt,conference]{IEEEtran}

\ifx\pdfoutput\undefined
\usepackage{graphicx}
\graphicspath{{eps_figs/}}
\else
\usepackage[pdftex]{graphicx}
\graphicspath{{pdf_figs/}}
\fi

\usepackage{latexsym,amssymb,amsmath,cite,ifthen,afterpage}

\newcommand{\Fig}[1]{Fig.~\ref{#1}}

\newcommand{\eqdef}{\stackrel{\scriptscriptstyle\bigtriangleup}{=} }
\newcommand{\smpl}[2]{#1^{(#2)}}

\newcommand{\R}{\mathbb{R}}
\newcommand{\C}{{\mathbb C}}
\newcommand{\calX}{\mathcal{X}}

\newcommand{\E}{\operatorname{E}}

\newcounter{examplecntr}
{\begin{trivlist}\small\item[]\refstepcounter{examplecntr}%
 {\bfseries Example~\theexamplecntr%
  \ifthenelse{\equal{#1}{}}{}{ (#1)}.
}}%
{\end{trivlist}}

\newcounter{theoremcntr}
{\begin{trivlist}\item[]\refstepcounter{theoremcntr}%
{\bfseries Theorem~\thetheoremcntr%
  \ifthenelse{\equal{#1}{}}{}{ (#1)}.
}}%
{\hfill$\Box$\end{trivlist}}

\begin{document}
\DeclareGraphicsExtensions{.pdf}

\title{Extending Monte Carlo Methods 
to Factor Graphs with Negative and Complex Factors}

\author{
\IEEEauthorblockN{Mehdi Molkaraie and Hans-Andrea Loeliger}
\IEEEauthorblockA{ETH Zurich\\
  Dept.\ of Information Technology \& Electrical Engineering\\ 
  8092 Z\"urich, Switzerland\\ 
  molkaraie@isi.ee.ethz.ch, loeliger@isi.ee.ethz.ch
}
}

\maketitle

\begin{abstract}
The partition function of a factor graph can sometimes be accurately estimated 
by Monte Carlo methods. 
In this paper, such methods are extended to factor graphs with negative and complex factors.


\end{abstract}

\section{Introduction}
\label{sec:Introduction}

Let $\calX_1, \calX_2, \ldots, \calX_N$ be finite sets,
let $\calX$ be the Cartesian product 
$\calX \eqdef \calX_1 \times \calX_2 \times \ldots \times \calX_N$,
and let $f$ be 
a function $f: \calX \rightarrow \C$.
We are interested in numerically computing the partition function
\begin{equation} \label{eqn:PartitionFunction}
Z_f \eqdef \sum_{x\in \calX} f(x)
\end{equation}
for cases where $N$ is large and 
$f$ has some useful factorization.

If $f$ has a cycle-free factor graph \cite{Lg:ifg2004}, 
the sum (\ref{eqn:PartitionFunction}) can be computed directly by sum-product message passing.
In this paper, however, we are interested in cases where no such factor graph is available 
(e.g., when the factor graph of $f$ is a rectangular grid as in \Fig{fig:2DGridT}).

In the important special case where $f$ is real-valued and nonnegative 
(i.e., $f(x) \geq 0$ for all $x\in \calX$), 
the probability mass function
\begin{equation} \label{eqn:positivepx}
p_f(x) \eqdef \frac{f(x)}{Z_f}
\end{equation}
is the basis of a variety of Monte Carlo methods for 
estimating~(\ref{eqn:PartitionFunction}), see~\cite{Neal:proinf1993r,MK:mct1998}. 
Such Monte Carlo algorithms have successfully been used, e.g., for 
the computation of information rates of source/channel models with 2-D memory 
\cite{LoMo:ITW2009,LoMo:2011}. 
Note that $p$ inherits factorizations
(and thus factor graphs) from $f$. 

In this paper, we extend these Monte Carlo methods
to the case where $f$ is real-valued (but not nonnegative) or complex.
The motivation for this extension is twofold. 
First, the Fourier transform of a function $f$ preserves the topology of the factor graph, 
but generally results in complex factors \cite{AY:2011,FV:2011}. 
Second, factor graphs of probability mass functions in quantum mechanics  
naturally involve complex functions \cite{LV:2012}. 
In both cases, computing quantities of the form (\ref{eqn:PartitionFunction}) 
is of supreme interest. 

In full generality, the computation of (\ref{eqn:PartitionFunction}) 
is, of course, intractable (already in the nonnegative real case), 
but good Monte Carlo algorithms may nonetheless work well 
for many cases of interest 
(as in the nonnegative real case).

The paper is structured as follows.
After introducing some notations in Section~\ref{sec:ProbPartialZ},
the proposed algorithms are described in Sections~\ref{sec:SecEstZf}
and~\ref{sec:EstcalX}.
The description focuses on the real case; the generalization to the 
complex case is outlined in Section~\ref{sec:ExtComplex}.
The proposed methods are illustrated by some numerical experiments 
in Section~\ref{sec:NumExpr}.

\section{Probabilities and Partial Partition Functions}
\label{sec:ProbPartialZ}

We begin the proposed generalization by defining
\begin{equation} \label{eqn:AbsPartitionFunction}
Z_{|f|} \eqdef \sum_{x\in\calX} |f(x)|
\end{equation}
and the probability mass function
\begin{equation} \label{eqn:Pdef}
p_{|f|}(x) \eqdef \frac{|f(x)|}{Z_{|f|}}
\end{equation}
which will replace (\ref{eqn:positivepx}) in the Monte Carlo algorithms.
Note that $p_{|f|}$ also inherits factorizations (and thus factor graphs) from $f$.

In the following, we restrict ourselves to the case 
where $f$ is real (but not nonnegative); the generalization to the complex case 
is indicated in Section~\ref{sec:ExtComplex}.

Let
\begin{IEEEeqnarray}{lCl} 
\calX^{+} &\eqdef& \{x\in\calX: f(x) > 0 \} \label{eqn:StatePos}\\
\calX^{-} &\eqdef& \{x\in\calX: f(x) < 0 \} \label{eqn:StateMin}\\
\calX^{0} &\eqdef& \{x\in\calX: f(x) = 0 \} \label{eqn:StateZer}
\end{IEEEeqnarray}
and thus
\begin{equation} \label{eqn:StateEq}
|\calX| = |\calX^{+}| + |\calX^{-}| + |\calX^{0}|.
\end{equation}
We then define the partial partition functions 
\begin{IEEEeqnarray}{lCl}
Z^{+}_f &\eqdef&  \sum_{x\in\calX^{+}} f(x) \label{eqn:ZfPSum}\\
Z^{-}_f &\eqdef& \sum_{x\in\calX^{-}} f(x) \label{eqn:ZfMSum}
\end{IEEEeqnarray}
%
and thus
\begin{IEEEeqnarray}{rCl} 
Z_f & = & Z^{+}_f + Z^{-}_f \label{eqn:Zplusminus} \\ 
Z_{|f|} & = & Z^{+}_f - Z^{-}_f. \label{eqn:ZAbsplusminus}
\end{IEEEeqnarray}


Finally, we define probability mass functions on $\calX^+$ and $\calX^-$ as follows:
\begin{equation} \label{eqn:pfplus}
p^{+}_f: \calX^+ \rightarrow \R: x \mapsto p^{+}_f(x) \eqdef \frac{f(x)}{Z^{+}_f}
\end{equation}
and
\begin{equation} \label{eqn:pfminus}
p^{-}_f: \calX^- \rightarrow \R: x \mapsto p^{-}_f(x) \eqdef \frac{f(x)}{Z^{-}_f}
\end{equation}
%

A simple, but key, insight is that sampling from $p^{+}_f$ or from $p^{-}_f$ 
can be reduced to sampling from $p_{|f|}$: 
since $p_{|f|}$ coincides with $p^{+}_f$ on $\calX^{+}$ (up to a scale factor), and 
with $p^{-}_f$ on $\calX^{-}$ (up to a scale factor), 
samples $\smpl{x}{1}, \smpl{x}{2}, \ldots$
from $p_{|f|}$ can be partitioned into samples from $p^{+}_f$ and from $p^{-}_f$
according to the sign of $f(\smpl{x}{\ell})$.
For example, samples from $p_{|f|}$, and thus both from $p^{+}_f$ and from $p^{-}_f$, 
may be drawn using tree-based Gibbs sampling as in \cite{LoMo:ITW2009,LoMo:2011,FF:f2t2004c}.

In this paper, we will now focus
on the computation/estimation of the partial partition functions 
$Z^{+}_f$ and $Z^{-}_f$ separately. 
If these estimates are sufficiently accurate, 
$Z_f$ can then be computed from (\ref{eqn:Zplusminus}). 
This approach is bound to fail, of course, if both $Z^{+}_f$ and $|Z^{-}_f|$ are large 
and their difference is small. 
However, this cancellation problem 
(which is well-known in quantum mechanics \cite{TW:05}) 
is beyond the scope of the present paper.

\section{Estimating $Z^{+}_f$ and $Z^{-}_f$}
\label{sec:SecEstZf}

We will now propose two different Monte Carlo methods 
to estimate the partial partition functions $Z^{+}_f$ and $Z^{-}_f$.
The first method uses uniform sampling and 
the second method uses samples from $p^{+}_f(x)$ and $p^{-}_f(x)$.
Both methods need the value of 
$|\calX^+|$ and $|\calX^-|$, the computation of which 
is addressed in Section~\ref{sec:EstcalX}.


\subsection{Uniform Sampling}
\label{sec:UnifZ}

\begin{enumerate}
\item Draw samples $x^{(1)}, x^{(2)}, \ldots, x^{(k)}, \ldots, x^{(K)}$ 
uniformly from $\calX^+$, and samples $x^{(1)}, x^{(2)}, \ldots, x^{(\ell)}, \ldots,
x^{(L)}$ 
uniformly from $\calX^-$.
\item Compute
\begin{IEEEeqnarray}{rCl}
\hat Z^+ &=& \frac{|\calX^{+}|}{K} \sum_{k = 1}^{K}f(x^{(k)}) \label{eqn:UnifP}\\
\hat Z^- &=& \frac{|\calX^{-}|}{L} \sum_{\ell = 1}^{L}f(x^{(\ell)}) \label{eqn:UnifN} 
\end{IEEEeqnarray}
\end{enumerate}
\hfill$\Box$

It is easily verified that $\E[\hat Z^+] = Z_f^{+}$ and 
$\E[\hat Z^-] = Z_f^-$. 

One way to draw samples uniformly from $\calX^+$ and/or $\calX^-$ 
is by drawing samples $x^{(1)}, x^{(2)}, \ldots, $ uniformly from $\calX$
and partitioning them according to the sign of $f(\smpl{x}{\ell})$.

\subsection{Ogata-Tanemura Method \cite{OgTa:eip1981,PoGo:sapf1997}}
\label{sec:OgataZ}


\begin{enumerate}
\item Draw samples $x^{(1)}, x^{(2)}, \ldots, x^{(k)}, \ldots, x^{(K)}$ 
from $\calX^+$ 
according to $p^{+}_f(x)$, as in (\ref{eqn:pfplus}), and 
samples $x^{(1)}, x^{(2)}, \ldots, x^{(\ell)}, \ldots, x^{(L)}$ from $\calX^-$ 
according to $p^{-}_f(x)$, as in (\ref{eqn:pfminus}).
\item Compute
\begin{IEEEeqnarray}{lCl}
\hat\Gamma^{+} &=& \frac{1}{K|\calX^{+}|} \sum_{k = 1}^{K}\frac{1}{f(x^{(k)})} 
\label{eqn:EstAbsZ2} \\
\hat\Gamma^{-} &=& \frac{1}{L|\calX^{-}|} \sum_{\ell = 1}^{L}\frac{1}{f(x^{(\ell)})}
\label{eqn:EstAbsZ3}
\end{IEEEeqnarray}
\end{enumerate}
\hfill$\Box$

It is easy to prove (see Appendix~A) that
$\E[\hat\Gamma^{+}] = \frac{1}{Z_f^{+}}$ and 
$\E[\hat\Gamma^-] = \frac{1}{Z_f^-}$. 

\section{Estimating $|\calX^{+}|$, $|\calX^{-}|$, and $|\calX^{0}|$}
\label{sec:EstcalX}

Again, we propose two different methods, one for uniform sampling and another for sampling from $p_{|f|}$.
In each case, the same samples as in Section~\ref{sec:SecEstZf} can be used.

\subsection{Uniform Sampling}
\label{sec:UnifX}

\begin{enumerate}
\item Draw samples $x^{(1)}, x^{(2)}, \ldots, x^{(k)}, \ldots, x^{(K)}$ uniformly from $\calX$.
\item Compute
\begin{IEEEeqnarray}{lCl} 
\xi^{+} &=& \frac{|\calX|}{K} \sum_{k = 1}^{K} [f(x^{k}) > 0] \label{eqn:EstXP} \\
\xi^{-} &=& \frac{|\calX|}{K} \sum_{k = 1}^{K} [f(x^{k}) < 0] \label{eqn:EstXN}\\
\xi^{0} &=& \frac{|\calX|}{K} \sum_{k = 1}^{K} [f(x^{k}) = 0] \label{eqn:EstXZ}
\end{IEEEeqnarray}
\vspace{-3.5ex}
\end{enumerate}
\hfill$\Box$

\noindent
In these equations, $[\cdot]$ denotes the Iverson bracket~\cite[p.\ 24]{GKP:89},
which evaluates to one if the condition in the bracket is satisfied 
and to zero otherwise.
It is easy to prove that 
$\E[\xi^{+}] = |\calX^{+}|$, $\E[\xi^{-}] = |\calX^{-}|$, and $\E[\xi^{0}] = |\calX^{0}|$.

\subsection{Sampling from $p_{|f|}$}
\label{sec:SysEquations}

We assume $|\calX^0| = 0$. 

\begin{enumerate}
\item Draw samples $x^{(1)}, x^{(2)}, \ldots, x^{(k)}, \ldots, x^{(K)}$ from $\calX$ 
according to $p_{|f|}$, as in~(\ref{eqn:Pdef}).
\item Compute
\begin{IEEEeqnarray}{rCl}
\hat\Lambda & = & \frac{1}{K} \sum_{k = 1}^{K} \frac{1}{f(x^{(k)})} \label{eqn:EstZ1} \\
\hat\Gamma & = & \frac{1}{K} \sum_{k = 1}^{K} \frac{1}{|f(x^{(k)})|}\label{eqn:EstAbsZ1} 
\end{IEEEeqnarray}
\end{enumerate}
\hfill$\Box$

It is not hard to prove (see Appendix~B) that 
$\E[\hat\Lambda] = \frac{|\calX^+| - |\calX^{-}|}{Z_{|f|}}$
and $\E[\hat\Gamma] = \frac{|\calX|}{Z_{|f|}}$.

Using~(\ref{eqn:StateEq}), 
we can then obtain estimates of  $|\calX^{+}|$ and $|\calX^{-}|$ from 
\begin{IEEEeqnarray}{rCl}
|\calX^{+}| + |\calX^{-}| & = & |\calX| \\
|\calX^+| - |\calX^{-}| & \approx & \frac{\hat\Lambda}{\hat\Gamma}|\calX|
\end{IEEEeqnarray}

\section{Extension to the Complex Case}
\label{sec:ExtComplex}

In (\ref{eqn:StatePos})--(\ref{eqn:ZfMSum}), 
we partitioned $\mathcal{X}$ and $Z_f$ according to 
the sign of $f(x)$. In the complex case, we allow more such bins, 
one for each possible argument (phase) of $f(x)$, as illustrated by 
the example in Section~\ref{sec:2DComplex}. The algorithms of 
Sections~\ref{sec:SecEstZf} and~\ref{sec:EstcalX} are easily generalized
 to this setting.
 
However, the computation of probabilities in factor graphs for quantum 
probabilities as in~\cite{{LV:2012}}, can actually be reduced to 
the real case as in 
Sections~\ref{sec:ProbPartialZ} and~\ref{sec:SecEstZf} 
(as will be detailed elsewhere).

\section{Numerical Experiments}
\label{sec:NumExpr}

In our numerical experiments,
we consider two-dimensional factor graphs of size
$N = m\times m$, with binary
variables $x_1, x_2, \ldots, x_N$, 
i.e., $\calX_1 = \calX_2 = \ldots = \calX_N = \{0,1\}$. 

We suppose 
$f: \{0,1\}^N \rightarrow \C$, 
and $f$ factors into 
\begin{equation} \label{eqn:Factorization}
f(x_1,\ldots,x_N) = \prod_{\text{$k,\ell$ adjacent}} \kappa(x_k,x_\ell)
\end{equation}
where the product runs over all adjacent pairs $(k,\ell)$.

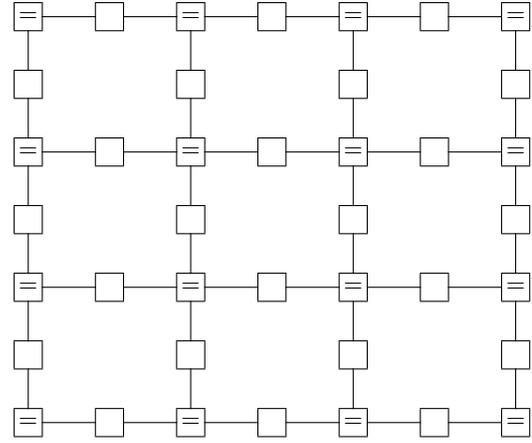
\begin{figure}
\setlength{\unitlength}{0.9mm}
\centering
\begin{picture}(76,64)(0,0)
\small
\put(0,60){\framebox(4,4){$=$}}
\put(4,62){\line(1,0){8}}
\put(12,60){\framebox(4,4){}}
\put(16,62){\line(1,0){8}}
\put(24,60){\framebox(4,4){$=$}}
\put(28,62){\line(1,0){8}}
\put(36,60){\framebox(4,4){}}
\put(40,62){\line(1,0){8}}
\put(48,60){\framebox(4,4){$=$}}
\put(52,62){\line(1,0){8}}
\put(60,60){\framebox(4,4){}}
\put(64,62){\line(1,0){8}}
\put(72,60){\framebox(4,4){$=$}}
\put(2,54){\line(0,1){6}}
\put(0,50){\framebox(4,4){}}
\put(2,50){\line(0,-1){6}}
\put(26,54){\line(0,1){6}}
\put(24,50){\framebox(4,4){}}
\put(26,50){\line(0,-1){6}}
\put(50,54){\line(0,1){6}}
\put(48,50){\framebox(4,4){}}
\put(50,50){\line(0,-1){6}}
\put(74,54){\line(0,1){6}}
\put(72,50){\framebox(4,4){}}
\put(74,50){\line(0,-1){6}}
\put(0,40){\framebox(4,4){$=$}}
\put(4,42){\line(1,0){8}}
\put(12,40){\framebox(4,4){}}
\put(16,42){\line(1,0){8}}
\put(24,40){\framebox(4,4){$=$}}
\put(28,42){\line(1,0){8}}
\put(36,40){\framebox(4,4){}}
\put(40,42){\line(1,0){8}}
\put(48,40){\framebox(4,4){$=$}}
\put(52,42){\line(1,0){8}}
\put(60,40){\framebox(4,4){}}
\put(64,42){\line(1,0){8}}
\put(72,40){\framebox(4,4){$=$}}
\put(2,34){\line(0,1){6}}
\put(0,30){\framebox(4,4){}}
\put(2,30){\line(0,-1){6}}
\put(26,34){\line(0,1){6}}
\put(24,30){\framebox(4,4){}}
\put(26,30){\line(0,-1){6}}
\put(50,34){\line(0,1){6}}
\put(48,30){\framebox(4,4){}}
\put(50,30){\line(0,-1){6}}
\put(74,34){\line(0,1){6}}
\put(72,30){\framebox(4,4){}}
\put(74,30){\line(0,-1){6}}
\put(0,20){\framebox(4,4){$=$}}
\put(4,22){\line(1,0){8}}
\put(12,20){\framebox(4,4){}}
\put(16,22){\line(1,0){8}}
\put(24,20){\framebox(4,4){$=$}}
\put(28,22){\line(1,0){8}}
\put(36,20){\framebox(4,4){}}
\put(40,22){\line(1,0){8}}
\put(48,20){\framebox(4,4){$=$}}
\put(52,22){\line(1,0){8}}
\put(60,20){\framebox(4,4){}}
\put(64,22){\line(1,0){8}}
\put(72,20){\framebox(4,4){$=$}}
\put(2,14){\line(0,1){6}}
\put(0,10){\framebox(4,4){}}
\put(2,10){\line(0,-1){6}}
\put(26,14){\line(0,1){6}}
\put(24,10){\framebox(4,4){}}
\put(26,10){\line(0,-1){6}}
\put(50,14){\line(0,1){6}}
\put(48,10){\framebox(4,4){}}
\put(50,10){\line(0,-1){6}}
\put(74,14){\line(0,1){6}}
\put(72,10){\framebox(4,4){}}
\put(74,10){\line(0,-1){6}}
\put(0,0){\framebox(4,4){$=$}}
\put(4,2){\line(1,0){8}}
\put(12,0){\framebox(4,4){}}
\put(16,2){\line(1,0){8}}
\put(24,0){\framebox(4,4){$=$}}
\put(28,2){\line(1,0){8}}
\put(36,0){\framebox(4,4){}}
\put(40,2){\line(1,0){8}}
\put(48,0){\framebox(4,4){$=$}}
\put(52,2){\line(1,0){8}}
\put(60,0){\framebox(4,4){}}
\put(64,2){\line(1,0){8}}
\put(72,0){\framebox(4,4){$=$}}
\end{picture}
\caption{\label{fig:2DGridT}
Forney factor graph with factors as in~(\ref{eqn:2DA}).}
\end{figure}

The corresponding Forney factor graph with factors as 
in~(\ref{eqn:Factorization}) is shown in \Fig{fig:2DGridT},
where the boxes labeled ``$=$'' are equality constraints \cite{Lg:ifg2004}. 

\begin{figure}[t]
\includegraphics[width=\linewidth, height = 5.76cm]{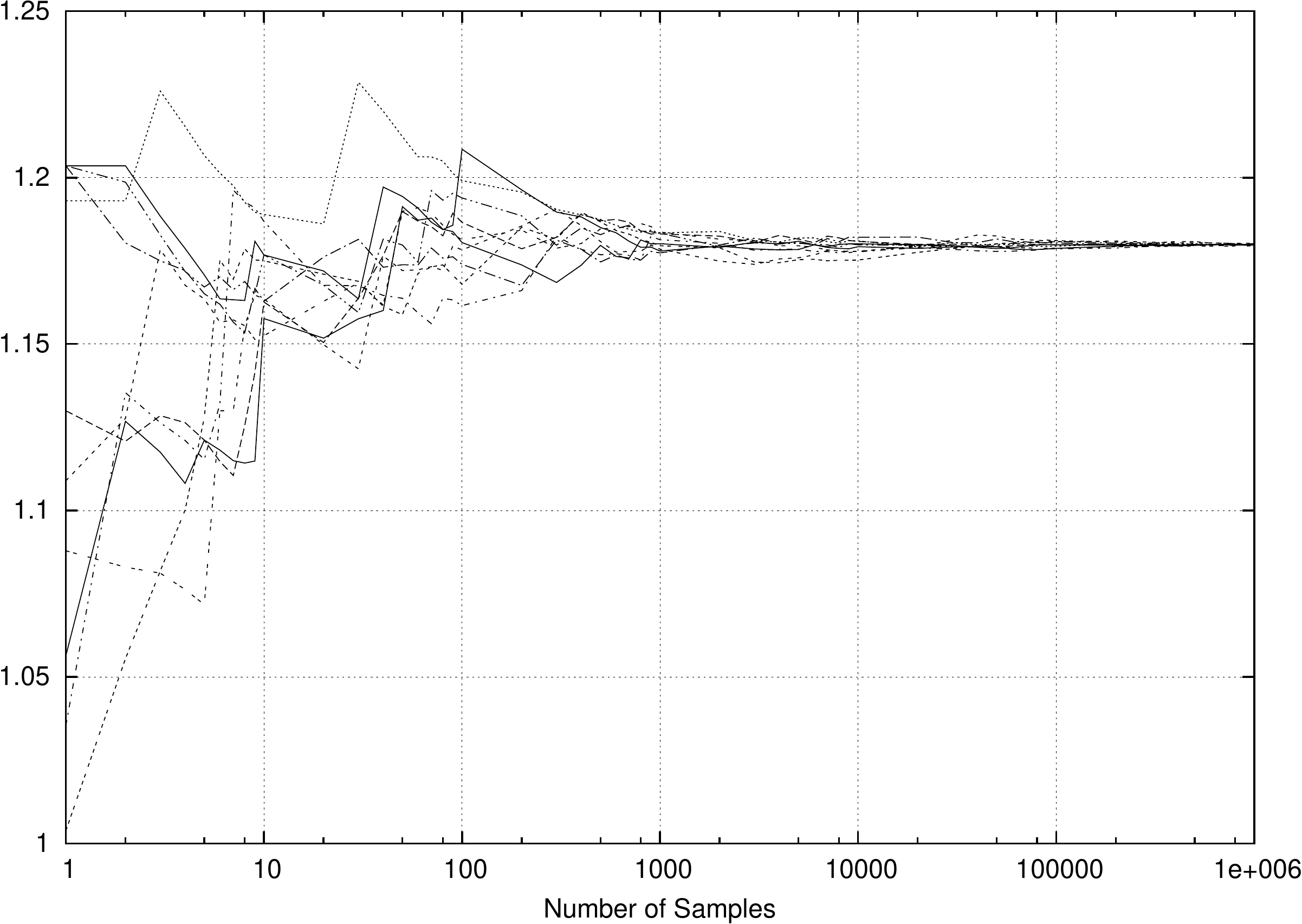}
\caption{\label{fig:Plot1}%
Estimated $\frac{1}{N}\log_2(Z_f^+)$ vs.\ the number of samples, 
for~$K = 10^5$, $N = 6\times 6$, and with factors as in~(\ref{eqn:NrgKernelEx}). 
The plot shows 10 sample paths each computed with estimator~(\ref{eqn:UnifP}).}
\end{figure}

\begin{figure}
\includegraphics[width=\linewidth, height = 5.76cm]{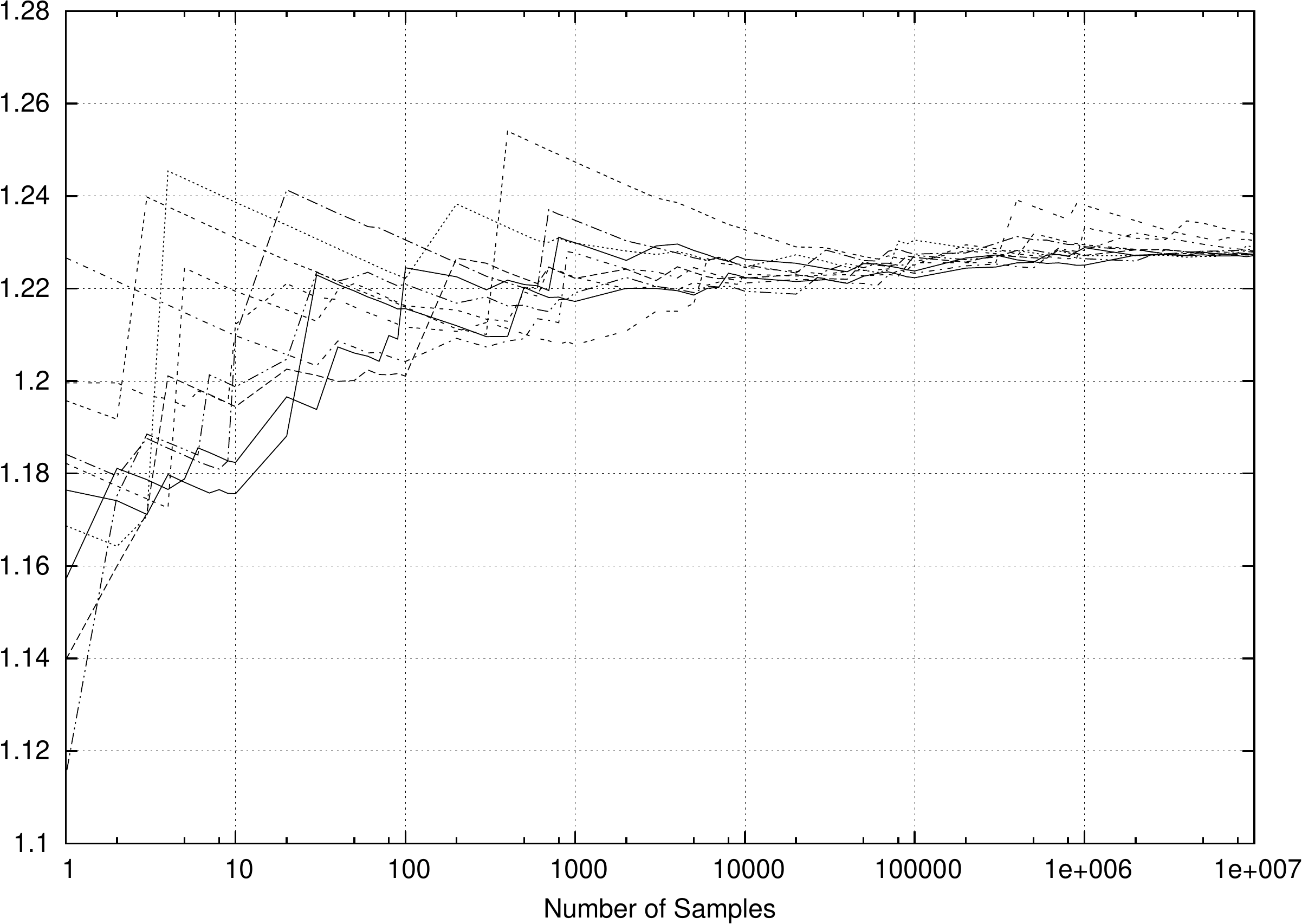}
\caption{\label{fig:Plot2}%
Same conditions as in Fig.~\ref{fig:Plot1}, but with $K = 10^7$ and $N = 14\times 14$.
}
\end{figure}
\begin{figure}[h]
\includegraphics[width=\linewidth, height = 5.76cm]{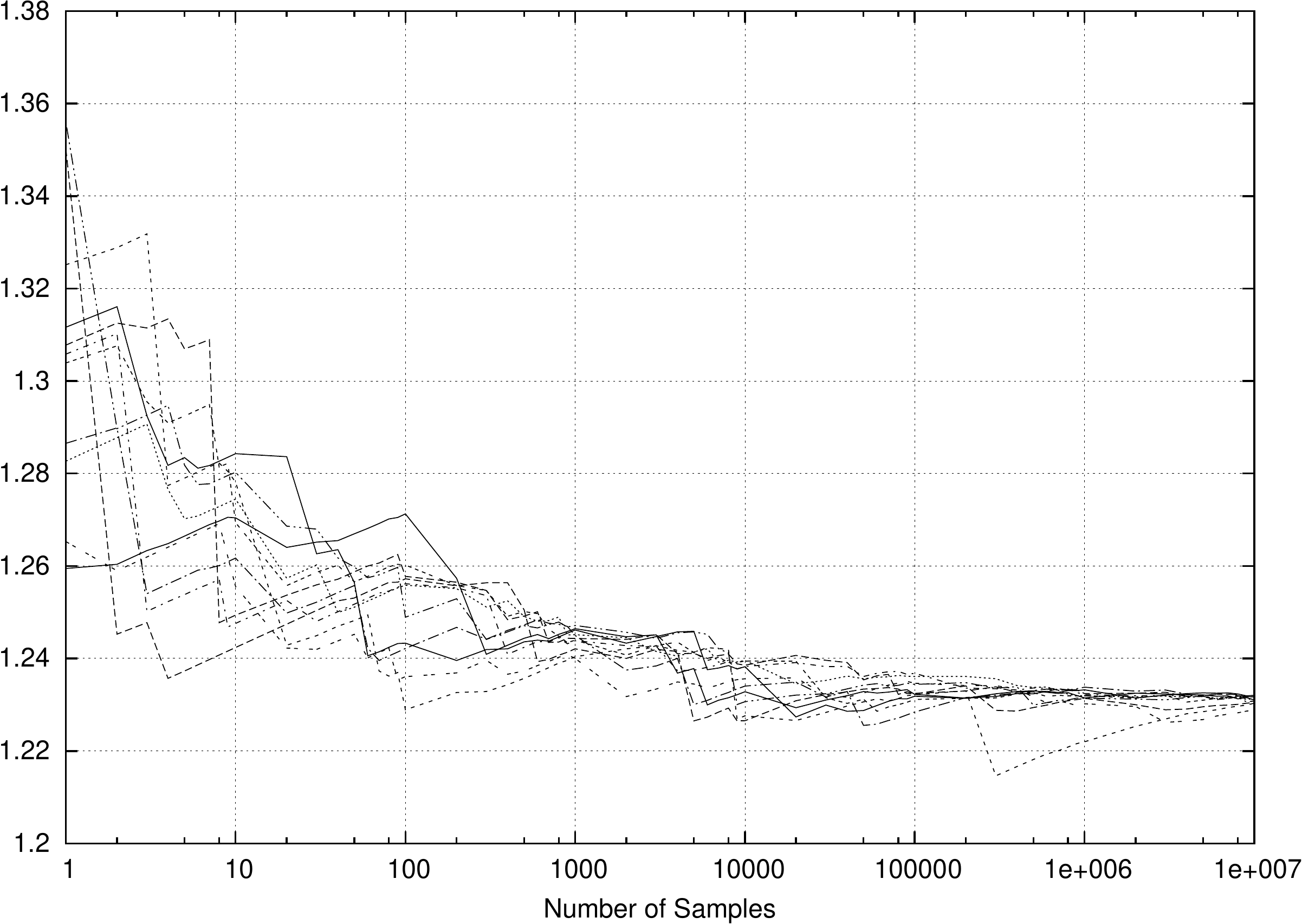}
\caption{\label{fig:Plot22}%
Estimated $\frac{1}{N}\log_2(Z_f^+)$ vs.\ the number of samples, 
for~$K = 10^7$, $N = 14\times 14$, and with factors as in~(\ref{eqn:NrgKernelEx}). 
The plot shows 10 sample paths each computed with estimator~(\ref{eqn:EstAbsZ2}).}
\end{figure}

\subsection{Two-Dimensional Model with Negative Factors}
\label{sec:2DNegative}

Let us consider a factor graph with factors as

\begin{equation} \label{eqn:NrgKernelEx}
\kappa(x_k,x_\ell) = \left\{ \begin{array}{ll}
     1.3, & \text{if $x_k = x_\ell = 0$} \\
     1, & \text{if $x_k = x_\ell = 1$} \\
     -1, & \text{otherwise}
  \end{array} \right.
\end{equation}

For this particular case, we prove in Appendix C that
$|\mathcal{X}^+|$ and $|\mathcal{X}^-|$ are analytically
available as
\begin{equation}
|\mathcal{X}^+| = |\mathcal{X}^-| = 2^{N-1}
\end{equation}

We estimate $Z_f^{+}$ using uniform sampling with
estimator~(\ref{eqn:UnifP}) 
of Section~\ref{sec:UnifZ}, and the
Oagata-Tanemura method with
estimator~(\ref{eqn:EstAbsZ2}) of Section~\ref{sec:OgataZ}.

\begin{figure}[t]
\includegraphics[width=\linewidth, height = 5.8cm]{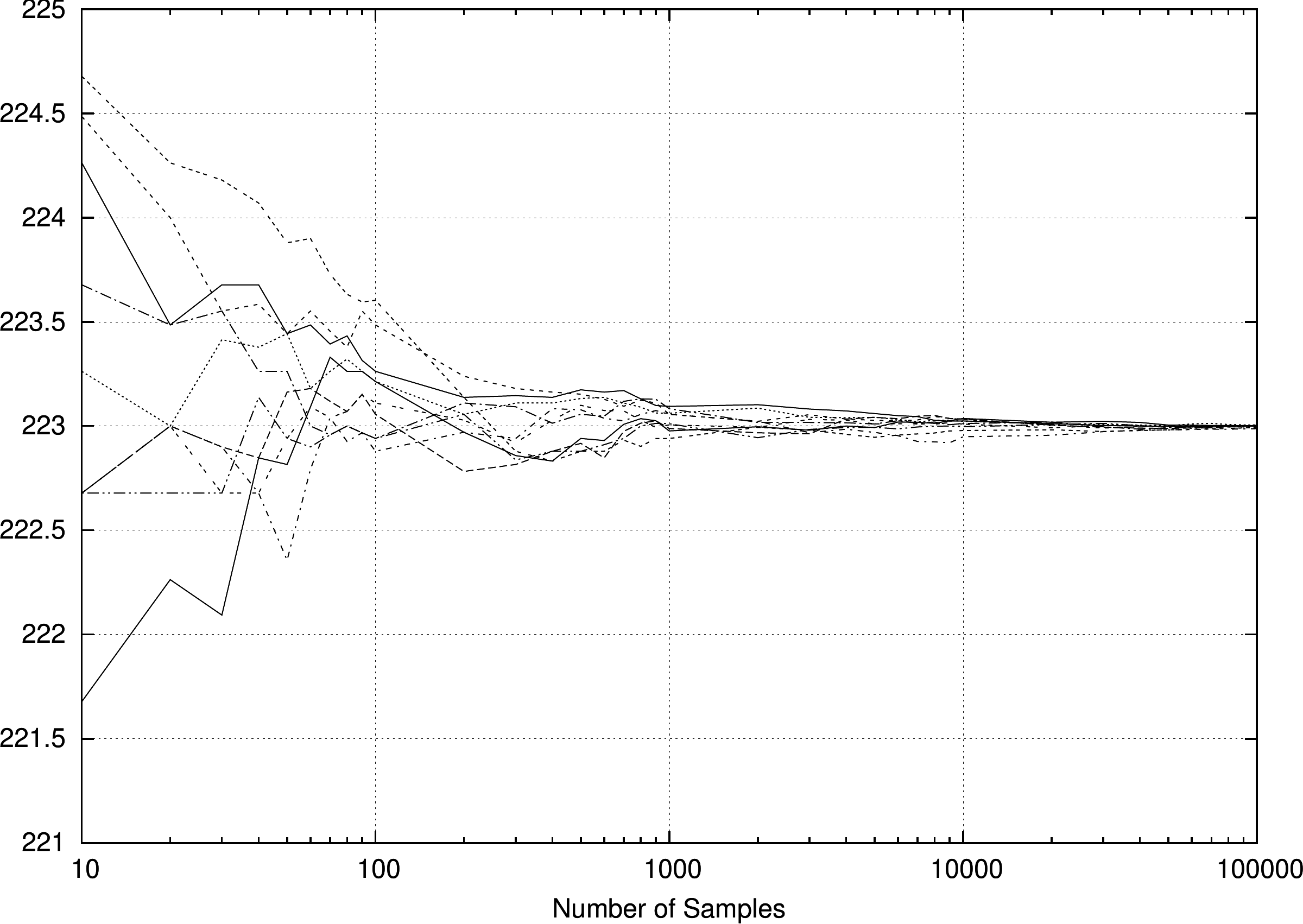}
\caption{\label{fig:Plot5}%
Estimated $\log_2(|\calX^+|)$ vs.\ the number of samples, for~$K = 10^5$,
$N = 15\times 15$, and with factors as in~(\ref{eqn:ImgKernelEx}). 
The plot shows 10 sample paths each computed with 
estimator~(\ref{eqn:EstXP}).}
\end{figure}


Some experimental results are shown in 
Figs.~\ref{fig:Plot1} through \ref{fig:Plot22}.
All figures refer to $f$ with factors as in~(\ref{eqn:NrgKernelEx}),
and show the quantity $\frac{1}{N}\log_2(Z_f^+)$
vs.\ the number of samples $K$. 

Figs.~\ref{fig:Plot1} and~\ref{fig:Plot2} show simulation 
results using uniform sampling and Fig.~\ref{fig:Plot22} 
using the Ogata-Tanemura method.

For $N = 6\times 6$, the 
estimated $\frac{1}{N}\log_2(Z_f^+)$ is about $1.18$,
and for 
$N = 14\times 14$, it is about $1.23$.
As discussed in Section~\ref{sec:SecEstZf}, 
$\frac{1}{N}\log_2(|Z_f^-|)$ can be computed analogously.

\subsection{Two-Dimensional Model with Complex Factors}
\label{sec:2DComplex}

We extend our numerical experiments to factor graphs with 
complex factors as
\begin{equation} \label{eqn:ImgKernelEx}
\kappa(x_k,x_\ell) = \left\{ \begin{array}{ll}
     1.5, & \text{if $x_k = x_\ell = 0$} \\
     i, & \text{if $x_k = x_\ell = 1$} \\
     1, & \text{otherwise}
  \end{array} \right.
\end{equation}
where $i$ is the unit imaginary number.

We define
\begin{IEEEeqnarray}{lCl}
\mathcal{X}^{(+i)} &\eqdef& \{x\in\calX: f(x) \in i\R_{>0}\} \label{eqn:StateImg}\\
\mathcal{X}^{(-i)} &\eqdef& \{x\in\calX: f(x) \in -i\R_{>0}\} \label{eqn:StateNImg}
\end{IEEEeqnarray}
where $\R_{>0} \eqdef \{x \in \R: x > 0\}$.
 
To estimate $|\mathcal{X}^+|$, $|\mathcal{X}^-|$, $|\mathcal{X}^{(i)}|$,
and $|\mathcal{X}^{(-i)}|$, we can apply uniform sampling 
of Section~\ref{sec:UnifX} by first drawing samples
$x^{(1)}, x^{(2)}, \ldots$ uniformly from $\mathcal{X}$,
and then using the samples in the relevant estimators
according to the value of $f(x^{(k)})$, e.g., in~(\ref{eqn:EstXP}) 
if $f(x^{(k)})$ is a real positive number.

For a factor graph of size 
$N = 15\times 15$, Fig.~\ref{fig:Plot5} shows the estimated 
$\log_2(|\calX^+|)$ vs.\ the number of samples $K$. We obtain 
$\log_2(|\calX^+|) \approx 223$.

\begin{figure}[t]
\includegraphics[width=\linewidth, height = 5.81cm]{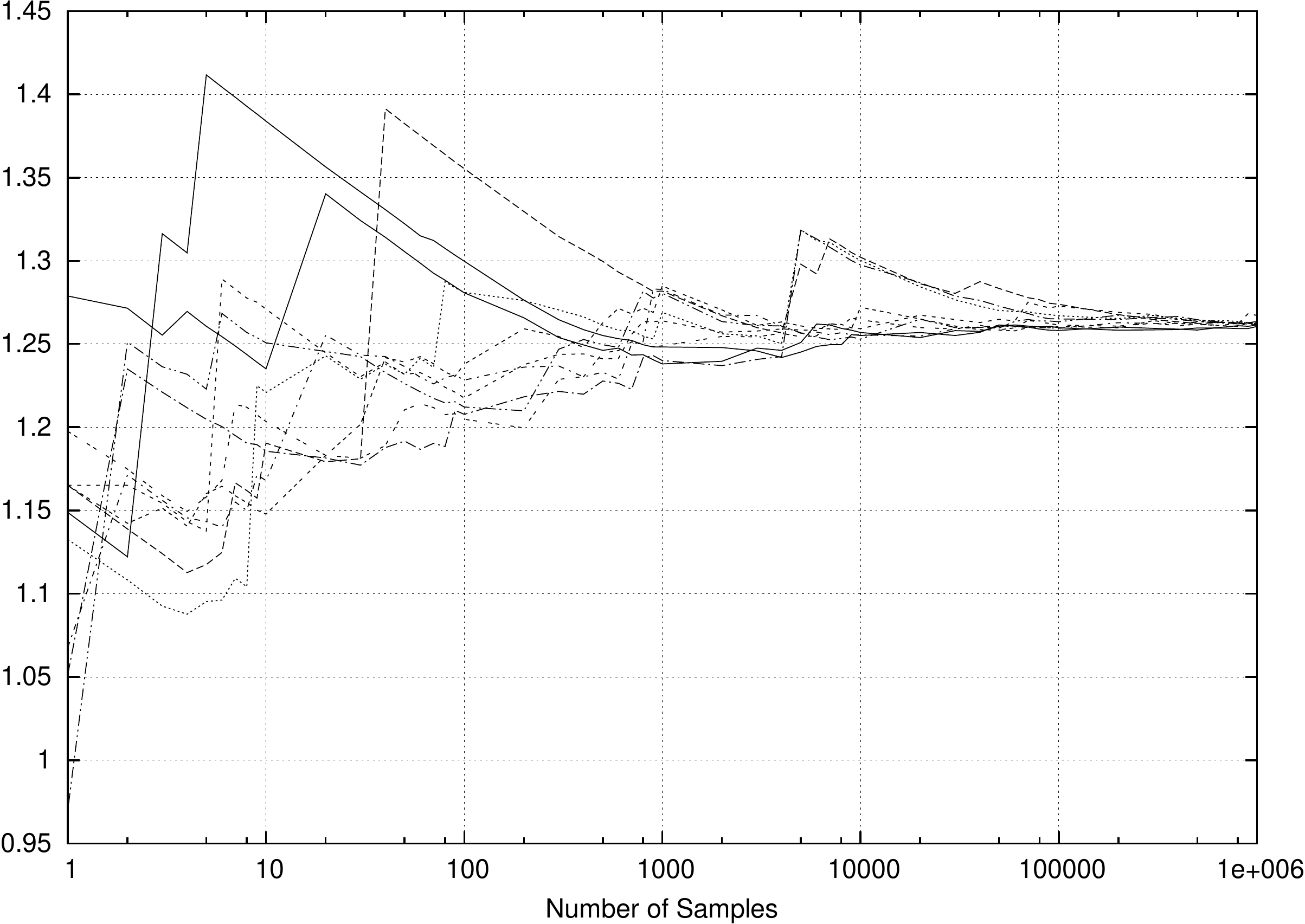}
\caption{\label{fig:Plot3}%
Estimated $\frac{1}{N}\log_2(Z_f^+)$ vs.\ the number of samples,
for~$K = 10^6$, $N = 6\times 6$, and with factors as in~(\ref{eqn:ImgKernelEx}). 
The plot shows 10 sample paths each computed with estimator~(\ref{eqn:UnifP}).}
\end{figure}

\begin{figure}[h]
\includegraphics[width=\linewidth, height = 5.81cm]{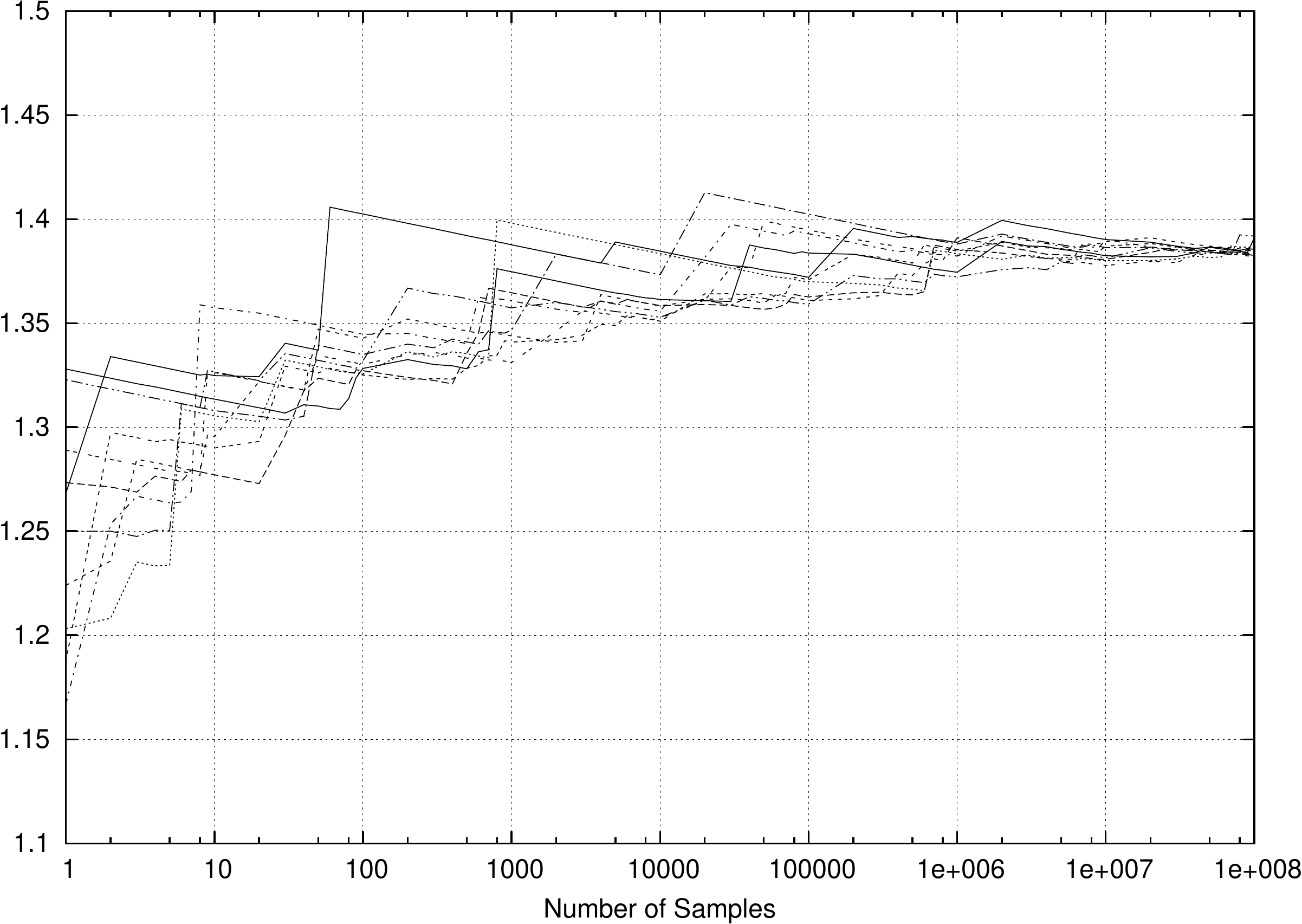}
\caption{\label{fig:Plot4}%
Estimated $\frac{1}{N}\log_2(Z_f^+)$ vs.\ the number of samples,
for~$K = 10^8$, $N = 15\times 15$, and with factors as in~(\ref{eqn:ImgKernelEx}). 
The plot shows 10 sample paths each computed with estimator~(\ref{eqn:EstAbsZ2}).}
\end{figure}

We again apply uniform sampling to estimate $Z_f^{+}$, 
see Section~\ref{sec:UnifZ}.
Some experimental results are shown in 
Figs.~\ref{fig:Plot3} and \ref{fig:Plot4}.
All figures refer to $f$ with factors as in~(\ref{eqn:ImgKernelEx}),
and show the quantity $\frac{1}{N}\log_2(Z_f^+)$
vs.\ the number of samples $K$. 

In Fig.~\ref{fig:Plot3}, we have
$N = 6\times 6$ and the estimated 
$\frac{1}{N}\log_2(Z_f^+)$ is about $1.26$.
In Fig.~\ref{fig:Plot4}, the estimated 
$\frac{1}{N}\log_2(Z_f^+)$ is about $1.38$ for
a factor graph of size $N = 15\times 15$.

\section{Conclusion}

We have shown that Monte Carlo methods as in \cite{LoMo:2011} 
can be extended to estimate the partition function 
of factor graphs with negative and complex factors. 
However, the cancellation problem
of partial partition functions as in (\ref{eqn:Zplusminus})
has not been addressed. 


\section*{Appendix A}
Suppose samples $x^{(1)}, x^{(2)}, \ldots, x^{(K)}$ are drawn
from $\calX^+$ according to $p^{+}_f(x)$, as in (\ref{eqn:pfplus}). 
We have
\begin{eqnarray} 
\E[\hat\Gamma^{+}] &=& \frac{1}{K|\calX^{+}|} \sum_{k = 1}^{K}\E\Big[\frac{1}{f(X^{(k)})}\Big] \\
& = & \frac{1}{K|\calX^{+}|} \sum_{k= 1}^{K}\sum_{x \in \calX^{+}}\frac{p^{+}_f(x)}{f(x)} \\
& = & \frac{1}{K|\calX^{+}|} \sum_{k = 1}^{K} \frac{|\calX^{+}|}{Z^{+}_f} \\
& = & \frac{1}{Z^{+}_f}
\end{eqnarray}
\hfill$\Box$

The proof
of $\E[\hat\Gamma^{-}] = \frac{1}{Z_f^{-}}$ follows along the same lines.

\section*{Appendix B}
Suppose samples $x^{(1)}, x^{(2)}, \ldots, x^{(K)}$ 
are drawn from $\calX$ 
according to $p_{|f|}$, as in~(\ref{eqn:Pdef}). We have
\begin{eqnarray} 
\E[\hat\Lambda] &=& \frac{1}{K} \sum_{k = 1}^{K}\E\Big[\frac{1}{f(X^{(k)})}\Big] \\
& = & \frac{1}{K} \sum_{k = 1}^{K}\sum_{x \in \calX}\frac{p_{|f|}(x)}{f(x)} \\
& = & \frac{1}{Z_{|f|}} \sum_{x \in \calX} \frac{|f(x)|}{f(x)} \\
& = & \frac{|\calX^+| - |\calX^{-}|}{Z_{|f|}}
\end{eqnarray}
\hfill$\Box$

The proof of $\E[\hat\Gamma] = \frac{|\calX|}{Z_{|f|}}$ follows along the same
lines.

\section*{Appendix C}

We consider a two-dimensional factor graph of size\linebreak
$N = m\times m$, where $m$ is finite and $m > 2$, with factors
\begin{equation} \label{eqn:2DA}
\kappa(x_k,x_\ell) = \left\{ \begin{array}{ll}
     a, & \text{if $x_k = x_{\ell}$} \\
     -a, & \text{otherwise}
  \end{array} \right.
\end{equation}
where $a \in \R$ and $a \ne 0$.

We use the normal factor graph duality theorem~\cite{AY:2011} to show 
that for this choice of factors, $Z_f$, as defined in~(\ref{eqn:PartitionFunction}),
is zero.

Consider the dual of the Forney factor graph 
with factors as in~(\ref{eqn:2DA}).
In the dual graph, the equality constraints are replaced by XOR factors,
and each factor~(\ref{eqn:2DA}) by its two-dimensional
Fourier transform which has the following form
\begin{equation} \label{eqn:2DZeroF}
\nu(\omega_k,\omega_\ell) = \left\{ \begin{array}{ll}
     4a, & \text{if $\omega_k = \omega_\ell = 1$} \\
     0, & \text{otherwise}
  \end{array} \right.
\end{equation}

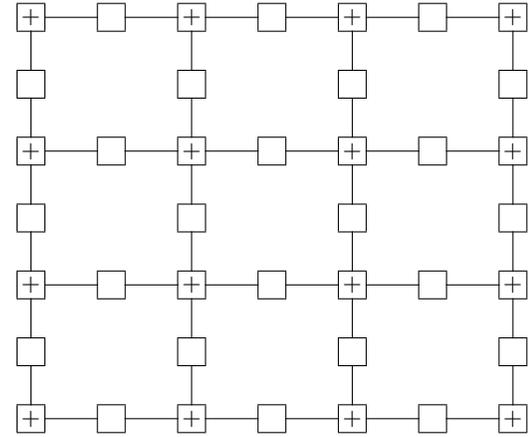
\begin{figure}
\setlength{\unitlength}{0.89mm}
\centering
\begin{picture}(76,64)(0,0)
\small
\put(0,60){\framebox(4,4){$+$}}
\put(4,62){\line(1,0){8}}
\put(12,60){\framebox(4,4){}}
\put(16,62){\line(1,0){8}}
\put(24,60){\framebox(4,4){$+$}}
\put(28,62){\line(1,0){8}}
\put(36,60){\framebox(4,4){}}
\put(40,62){\line(1,0){8}}
\put(48,60){\framebox(4,4){$+$}}
\put(52,62){\line(1,0){8}}
\put(60,60){\framebox(4,4){}}
\put(64,62){\line(1,0){8}}
\put(72,60){\framebox(4,4){$+$}}
\put(2,54){\line(0,1){6}}
\put(0,50){\framebox(4,4){}}
\put(2,50){\line(0,-1){6}}
\put(26,54){\line(0,1){6}}
\put(24,50){\framebox(4,4){}}
\put(26,50){\line(0,-1){6}}
\put(50,54){\line(0,1){6}}
\put(48,50){\framebox(4,4){}}
\put(50,50){\line(0,-1){6}}
\put(74,54){\line(0,1){6}}
\put(72,50){\framebox(4,4){}}
\put(74,50){\line(0,-1){6}}
\put(0,40){\framebox(4,4){$+$}}
\put(4,42){\line(1,0){8}}
\put(12,40){\framebox(4,4){}}
\put(16,42){\line(1,0){8}}
\put(24,40){\framebox(4,4){$+$}}
\put(28,42){\line(1,0){8}}
\put(36,40){\framebox(4,4){}}
\put(40,42){\line(1,0){8}}
\put(48,40){\framebox(4,4){$+$}}
\put(52,42){\line(1,0){8}}
\put(60,40){\framebox(4,4){}}
\put(64,42){\line(1,0){8}}
\put(72,40){\framebox(4,4){$+$}}
\put(2,34){\line(0,1){6}}
\put(0,30){\framebox(4,4){}}
\put(2,30){\line(0,-1){6}}
\put(26,34){\line(0,1){6}}
\put(24,30){\framebox(4,4){}}
\put(26,30){\line(0,-1){6}}
\put(50,34){\line(0,1){6}}
\put(48,30){\framebox(4,4){}}
\put(50,30){\line(0,-1){6}}
\put(74,34){\line(0,1){6}}
\put(72,30){\framebox(4,4){}}
\put(74,30){\line(0,-1){6}}
\put(0,20){\framebox(4,4){$+$}}
\put(4,22){\line(1,0){8}}
\put(12,20){\framebox(4,4){}}
\put(16,22){\line(1,0){8}}
\put(24,20){\framebox(4,4){$+$}}
\put(28,22){\line(1,0){8}}
\put(36,20){\framebox(4,4){}}
\put(40,22){\line(1,0){8}}
\put(48,20){\framebox(4,4){$+$}}
\put(52,22){\line(1,0){8}}
\put(60,20){\framebox(4,4){}}
\put(64,22){\line(1,0){8}}
\put(72,20){\framebox(4,4){$+$}}
\put(2,14){\line(0,1){6}}
\put(0,10){\framebox(4,4){}}
\put(2,10){\line(0,-1){6}}
\put(26,14){\line(0,1){6}}
\put(24,10){\framebox(4,4){}}
\put(26,10){\line(0,-1){6}}
\put(50,14){\line(0,1){6}}
\put(48,10){\framebox(4,4){}}
\put(50,10){\line(0,-1){6}}
\put(74,14){\line(0,1){6}}
\put(72,10){\framebox(4,4){}}
\put(74,10){\line(0,-1){6}}
\put(0,0){\framebox(4,4){$+$}}
\put(4,2){\line(1,0){8}}
\put(12,0){\framebox(4,4){}}
\put(16,2){\line(1,0){8}}
\put(24,0){\framebox(4,4){$+$}}
\put(28,2){\line(1,0){8}}
\put(36,0){\framebox(4,4){}}
\put(40,2){\line(1,0){8}}
\put(48,0){\framebox(4,4){$+$}}
\put(52,2){\line(1,0){8}}
\put(60,0){\framebox(4,4){}}
\put(64,2){\line(1,0){8}}
\put(72,0){\framebox(4,4){$+$}}
\end{picture}
\caption{\label{fig:2DGridF}
The dual Forney factor graph where unlabeled boxes represent
factors as in~(\ref{eqn:2DZeroF}).}
\end{figure}

The corresponding Forney factor graph of the dual graph
is shown in \Fig{fig:2DGridF},
where the unlabeled boxes represent factors as in~(\ref{eqn:2DZeroF}).

Let us denote the partition function of the dual graph by 
$Z_d$. 
Note that, each factor $\nu(\omega_k,\omega_\ell)$ is non-zero 
if $\omega_k = \omega_\ell = 1$. Therefore, only the all-ones pattern might
have a non-zero contribution to $Z_d$. 
But this pattern does not satisfy the XOR 
factors of degree three in the dual graph, therefore
$Z_d = 0$. Using the normal factor graph duality theorem
\cite[Theorem 2]{AY:2011}, \cite{FV:2011},
we conclude that $Z_f = 0$. 
Therefore, using~(\ref{eqn:Zplusminus}) and~(\ref{eqn:ZAbsplusminus}), 
we obtain
\begin{equation}
\label{eqn:ZPNExact}
Z_f^{+} = -Z_f^{-} = \frac{Z_{|f|}}{2}
\end{equation}

Putting $a = 1$ (or $a = -1$), we have
\begin{equation}
Z_f^{+} = -Z_f^{-} = 2^{N-1}
\end{equation}
and hence the following
\begin{IEEEeqnarray}{rCl}
 Z_f^{+} &=& |\mathcal{X}^{+}|\\
 -Z_f^{-} &=& |\mathcal{X}^{-}|
\end{IEEEeqnarray}

Thus
\begin{equation}
\label{eqn:CalXExact}
|\mathcal{X}^{+}| = |\mathcal{X}^{-}| = 2^{N-1}.
\end{equation}

Note that for $m = 2$, we have
$|\mathcal{X}^{+}| = |\mathcal{X}|$, $Z_f^{+} = Z_f$,
and $Z_f^{-} = 0$.
Finally,
note that, we can still show $Z_f = 0$ if $a \in \C$.

\section*{Acknowledgement}
The first author would like to thank Radford Neal, Ruslan Salakhutdinov,
and Neal Madras for helpful discussions. The 
authors would also like to thank Pascal Vontobel for his 
helpful comments on an earlier draft of this paper.

\newcommand{\IT}{IEEE Trans.\ Inf.\ Theory}
\newcommand{\CASI}{IEEE Trans.\ Circuits \& Systems~I}
\newcommand{\COM}{IEEE Trans.\ Comm.}
\newcommand{\COMLet}{IEEE Commun.\ Lett.}
\newcommand{\COMMag}{IEEE Communications Mag.}
\newcommand{\ETT}{Europ.\ Trans.\ Telecomm.}
\newcommand{\SPMag}{IEEE Signal Proc.\ Mag.}
\newcommand{\ProcIEEE}{Proceedings of the IEEE}


\begin{thebibliography}{10}

\bibitem{Lg:ifg2004}
H.-A.\ Loeliger,
``An introduction to factor graphs,''
\emph{\SPMag,} Jan.\ 2004, pp.~28--41.

\bibitem{Neal:proinf1993r}
R.~M.~Neal,
\emph{Probabilistic Inference Using Markov Chain Monte Carlo Methods,}
Techn.\ Report CRG-TR-93-1, Dept.\ Comp.\ Science, Univ.\ of Toronto, Sept.\ 1993.


\bibitem{MK:mct1998}
D.~J.~C.~MacKay,
``Introduction to Monte Carlo methods,'' in 
\emph{Learning in Graphical Models,} M.~I.~Jordan, ed., 
Kluwer Academic Press, 1998, pp.~175--204. 


\bibitem{LoMo:ITW2009}
H.-A.~Loeliger and M.~Molkaraie,
``Estimating the partition function of \mbox{2-D} fields
and the capacity of constrained noiseless \mbox{2-D} channels using 
tree-based Gibbs sampling,''
\emph{Proc.\ 2009 IEEE Information Theory Workshop,}
Taormina, Italy, October 11--16, pp.~228--232.

\bibitem{LoMo:2011}
M.~Molkaraie and H.-A.~Loeliger,
``Monte Carlo algorithms for the partition function and information rates 
of two-dimensional channels,''
arXiv:1105.5542, 2011.

\bibitem{AY:2011}
A.~Al-Bashabsheh and Y.~Mao,
``Normal factor graphs and holographic transformations,"
\emph{\IT}, vol.~57, no.~2, pp.~752--763, Feb.\ 2011.

\bibitem{FV:2011}
G.~D.~Forney, Jr.\ and P.~O.~Vontobel,
``Partition functions of normal factor graphs,"
\emph{Proc.\ Information Theory and Applications Workshop}, UCSD, 
CA, USA, Feb.\ 2011.

\bibitem{LV:2012}
H.-A.~Loeliger and P.~O.~Vontobel,
``A factor-graph representation of probabilities in quantum mechanics,''
\emph{IEEE Int.\ Symp.\ on Information Theory,}
Cambridge, USA, July 1--6, 2012.

\bibitem{FF:f2t2004c}
F.~Hamze and N.~de Freitas,
``From fields to trees,''
\emph{Proc.\ Conf.\ on Uncertainty in Artificial Intelligence,} Banff, July 2004.


\bibitem{TW:05}
M.~Troyer and U.-J.~Wiese,
``Computational complexity and fundamental limitations to fermionic quantum Monte Carlo simulations,''
\emph{Phys. Rev. Lett,} vol.~94, May 2005.


\bibitem{PoGo:sapf1997}
G.~Potamianos and J.~Goutsias,
``Stochastic approximation algorithms
for partition function estimation of Gibbs random fields,''
\emph{\IT,} vol.~43, pp.~1984--1965, Nov.~1997.



\bibitem{OgTa:eip1981}
Y.~Ogata and M.~Tanemura,
``Estimation of interaction potentials of spatial point patterns
through the maximum likelihood procedure,''
\emph{Ann.\ Inst.\ Statist.\ Math.,} vol.~33, pp.~315--338, 1981.

\bibitem{GKP:89}
R.~L.~Graham, D.~E.~Knuth, and O.~Patashnik
\emph{Concrete Mathematics: A Foundation for Computer Science.}
Addison-Wesley, 1989.






\end{thebibliography}
\end{document}